\documentclass{article}
\usepackage[english]{babel}
\usepackage[a4paper, top=2cm, bottom=2cm, left=3cm, right=3cm, marginparwidth=1.75cm]{geometry}

\usepackage{amssymb}
\usepackage{siunitx}
\usepackage{hyperref}
\usepackage{cleveref}
\usepackage[utf8]{inputenc}
\usepackage[right]{lineno}
\usepackage{csquotes}
\usepackage{booktabs}
\usepackage{longtable}
\usepackage{adjustbox}
\usepackage{array}
\usepackage{url}
\usepackage{titlesec}
\usepackage{authblk}
\usepackage{xcolor}
\titleformat{\subsection}
  {\mdseries\itshape\large} 
  {\thesubsection}{1em}{} 

\crefformat{figure}{#2Figure~#1#3}
\Crefformat{figure}{#2Figure~#1#3}
\crefformat{table}{#2Table~#1#3}
\Crefformat{table}{#2Table~#1#3}
\crefformat{section}{#2Section~#1#3}
\Crefformat{section}{#2Section~#1#3}

\author[1, 2]{Henry Bloss$^{*\mathsection}$}
\author[1]{Brandon Kriesten$^\dagger$}
\author[1]{T. J. Hobbs$^\ddagger$}
\affil[1]{High Energy Physics Division, Argonne National Laboratory, Lemont, IL 60439}
\affil[2]{Department of Physics and Astronomy, University of Notre Dame, Notre Dame, IN 46556}

\title{Quantum entropy and QCD factorization for low-$Q^2$ $\nu$DIS \\ {\large Contribution to the 25th International Workshop on Neutrinos from Accelerators}}

\begin{document}
\maketitle
$\hspace{-5mm} ^*$DOE SULI student, Summer 2024. Preprint, ANL-193386. \\
$\hspace{-5mm} ^\mathsection$\href{hbloss@nd.edu}{hbloss@nd.edu} \ \ \ \ $^\dagger$\href{bkriesten@anl.gov}{bkriesten@anl.gov} \ \ \ \ $^\ddagger$\href{tim@anl.gov}{tim@anl.gov} \\

\begin{abstract}
Deeply inelastic scattering (DIS) is an essential process for exploring the structure of visible matter and testing the standard model.
At the same time, the theoretical interpretation of DIS measurements depends on QCD factorization theorems whose validity deteriorates at the lower values of $Q^2$ and $W^2$ typical of neutrino DIS in accelerator-based oscillation searches.
For this reason, progress in understanding the origin and limits of QCD factorization is invaluable to the accuracy and precision of predictions for these upcoming neutrino experiments.
In these short proceedings, we introduce a novel approach based on the quantum entropy associated with continuous distributions in QCD, using
it to characterize the limits of factorization theorems relevant for the description of neutrino DIS.
This work suggests an additional avenue for dissecting factorization-breaking dynamics through the quantum entropy, which could also play a role in quantum simulations of related systems.
\end{abstract}

\section{Introduction}
Deeply inelastic scattering (DIS) is expected to represent a significant contribution to events recorded at the upcoming DUNE science runs, particularly at the multi-GeV $E_\nu$ energies~\cite{Ruso:2022qes} for which the neutrino flux distribution has a long tail.
Neutrino DIS belongs within a larger landscape of neutrino-scattering measurements with the potential to test the standard model (SM), including with respect to the possible presence of non-standard interactions (NSI) \cite{Falkowski:2021bkq}.
At higher energies, neutrino DIS is dominated by continuum DIS, wherein interactions are effectively characterized by scattering from individual partons in the target and the perturbative QCD expansion of the associated DIS structure function is generally stable; for this regime, theoretical predictions~\cite{Xie:2023suk} are now approaching next-to-next-to-next-to-leading order (N3LO) accuracy in the $\alpha_s$ series.
The foundation for these calculations is the formalism of QCD factorization~\cite{Collins:2011zzd}, embodied by {\it factorization theorems} of the type
\begin{equation}
\label{eq:fact}
F(x, Q^2) = \sum_i \sum_j \bigg\{ C_{i,j} \otimes \Phi_j \bigg\}(x, Q^2)\, +\, \mathcal{O}\left({M^2 \over Q^2}\right)\ ,
\end{equation}
where the short-distance Wilson coefficient functions, $C_{i,j}$, are perturbatively calculable, $\Phi_j$ contain quantum correlation functions like the nucleon or nuclear parton distribution functions (PDFs), and the indices $(i,j)$ are parton-flavor labels; in charge-current neutrino DIS, the $C_{i,j}$ do not vanish for $i\! \neq\! j$ due to the flavor-changing nature of the parton-level interaction.
Of particular note are the power-suppressed, $\sim\!\! 1/Q^2$ contributions which correct the all-orders factorized expression represented by the first term of Eq.~(\ref{eq:fact}).

\begin{figure}[ht]
  \centering
  \makebox[\textwidth][c]{\includegraphics[width=0.5\textwidth]{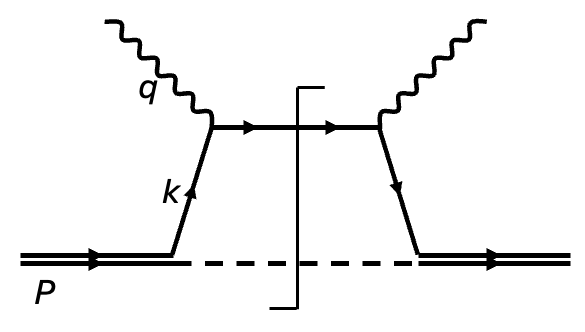}}%
  \caption{In the spectator model of Ref.~\cite{Moffat:2017sha}, the DIS interaction can be computed exactly without any factorization assumptions in the form of the handbag shown here; additional gauge invariance-preserving diagrams are not shown here for simplicity.  }
  \label{fig-1}
\end{figure}

A longstanding issue in the theoretical description of DIS has therefore been the validity of factorization theorems like Eq.~(\ref{eq:fact}), especially given the presence of the $\mathcal{O}(M^2/Q^2)$ factorization-breaking effects; in principle, these arise through a variety of kinematical and dynamical mechanisms, including the effects of non-leading twist (twist-$4$ for neutrino scattering from unpolarized targets) or sub-leading effects associated with finite transverse momentum, $k_T$, in the DIS interaction.
The theoretical challenges associated with describing the kinematical region of comparatively low $Q^2$ and $W^2$ has motivated phenomenological models~\cite{Bodek:2002ps} which extrapolate structure functions from the continuum DIS region down into the more intermediate regime of shallow scattering and beyond.
More broadly, ambiguities in the onset and exact dependence on $x$ and $Q^2$ of effects which break factorization are such that theoretical models which allow DIS processes to be computed both with and without the explicit assumption of factorization are valuable. An example of such a QCD-inspired model is that of Moffat {\it et al.}~\cite{Moffat:2017sha}, elements of which we deploy in this analysis. In particular, Ref.~\cite{Moffat:2017sha} formulated the structure of the interacting nucleon as a struck quark and recoiling spectator diquark as depicted in the diagram of Fig.~\ref{fig-1}, in which the essential degrees-of-freedom of the model are the $\mathcal{O}(100\,\mathrm{MeV})$ masses of these constituents.
The advantage of this approach is that it is then possible to compute DIS structure functions consistently and exactly from the appropriate diagrams of the type shown in Fig.~\ref{fig-1}, in addition to parallel calculations which explicitly assume factorization along the lines of Eq.~(\ref{eq:fact}).

In these short proceedings, we provide a preview demonstration of how notions of quantum entropy may be applied to DIS within such a model; we note that the results shown here will be presented in a longer study~\cite{2025QuantumEntropy} to appear soon. In particular, we define a suitable entropy of continuous distributions as provided by a model like that of Ref.~\cite{Moffat:2017sha}, and monitor how the factorization-breaking effects which enter the model imprint themselves on the entropy.
Our results suggest a relation between quantum mechanical entropies and the dynamics in a simplified model that captures aspects of QCD. We briefly introduce the theoretical motivation for the entropy definition(s) we use in Sec.~\ref{sec:entropy} before highlighting an example of the associated numerical results in Sec.~\ref{sec:results}; we indicate next steps in Sec.~\ref{sec:conc}.

\section{Localization, quantum entropy, and QCD factorization}
\label{sec:entropy}
There are compelling reasons~\cite{Aidala} to expect that aspects of high-energy interactions involving QCD bound states may admit a description in terms of decoherence and entanglement entropy.
For example, the validity of Eq.~(\ref{eq:fact}) is based effectively on a notion of decoherence between the dynamical effects contributing to the long-distance physics inherent to the PDFs ($\Phi_j$) and the short-distance interactions of the partonic cross section ($C_{i,j}$). By this logic, residual correlations between the factorized sub-processes of the DIS interaction would register as a form of entanglement which could leave a signature on related definitions of the entropy while simultaneously spoiling factorization.
In the context of models like Ref.~\cite{Moffat:2017sha}, these effects enter as, {\it e.g.}, $\sim\! k^2_T / Q^2$ deviations which can be computed precisely.
Related to these observations, it has also been shown~\cite{Hagiwara:2018sha} as possible to formulate entropies of entanglement from quantities related to the multi-dimensional wave function of the proton, including the Wehrl entropy~\cite{WEHRL1979353} as assessed on the Wigner function.

Using the model of Ref.~\cite{Moffat:2017sha}, we compute $k_T$-dependent structure functions obeying
\begin{equation}
\label{eq:norm}
    F_1(x,Q^2) = {1 \over (2\pi)^2}\int d^{2}\mathbf{k}_{T}\ f_1(x, k_T, Q^2)\ ,
\end{equation}
such that the unintegrated structure function, $f_1(x, k_T, Q^2)$, permits an interpretation as a distribution. In the discussion below, we normalize $f_1(x, k_T, Q^2)$ by hand according to Eq.~(\ref{eq:norm}).
On this basis, we define a {\it differential entropy} associated with the $k_T$ distribution~\cite{Hagiwara:2018sha} over the model structure function, $F_1$; namely,
\begin{equation}
\label{eq:entropy_int}
    H_{F_1}(x,Q^2) = - {1 \over (2\pi)^2} \int d^{2}\mathbf{k}_{T}\ f_1(x, k_T, Q^2)\, \ln \left[f_1(x, k_T, Q^2)\right]\ ;
\end{equation}
to gain insights into this quantity, we explore its $k_T$-dependent integrand --- specifically,
\begin{equation}
\label{eq:entropy}
    \mathcal{H}_{F_1}(x, k_T, Q^2)\ \equiv\ -k_T f_1(x, k_T, Q^2)\, \ln \left[ f_1(x, k_T, Q^2) \right]\ .
\end{equation}
While care is needed in the interpretation of entropies defined over continuous distributions as above, the differential entropy is mathematically connected to notions of localization as reflected in the shape of the underlying densities --- for instance, more localized systems generally produce sharply peaked distributions with correspondingly strongly negative differential entropies. As Eq.~(\ref{eq:entropy}) emerges from an inherently quantum mechanical object, these indications of localization may be connected in a model context to the decoherence of the associated wave function.

\section{Preliminary results}
\label{sec:results}
While the model calculation itself is essentially for generic DIS, we compute our results at kinematical points of relevance to neutrino scattering --- at the threshold of the soft DIS region, which is conventionally taken to occur near $Q\! \sim\! 1$ GeV$^2$ and $W\!\lesssim\! 2$ GeV. In particular, we select a virtuality modestly larger than this transition scale, taking $Q\! =\!2$ GeV, and compute the quantum entropies defined above for $x\! =\! 0.6$. Using the well-known expression for the DIS invariant mass, $W^2 = M^2 + Q^2/x -Q^2$, these kinematics correspond to $(Q, W) = (2, 1.9)\,$GeV. Similarly, in using the model of Ref.~\cite{Moffat:2017sha}, we select typical values for the internal parameters, with $m_q = 0.3,\, 0.5$ GeV for the constituent quark, and $m_s = 0.67\!-\!0.75$ GeV for the scalar spectator.

With these choices, we may compute the differential entropies summarized in Sec.~\ref{sec:entropy} above, evaluating both the $k_T$-(un)integrated objects of Eqs.~(\ref{eq:entropy}) and (\ref{eq:entropy}), respectively. These quantities can be evaluated for both the exact and factorized scenarios of the model, such that we quantify the impact of factorization-breaking effects on the entropy associated with the $k_T$ distributions.

\begin{figure}[ht]
  \vspace{-0.3cm}
  \centering
  \makebox[\textwidth][c]
  {\includegraphics[width=\textwidth]{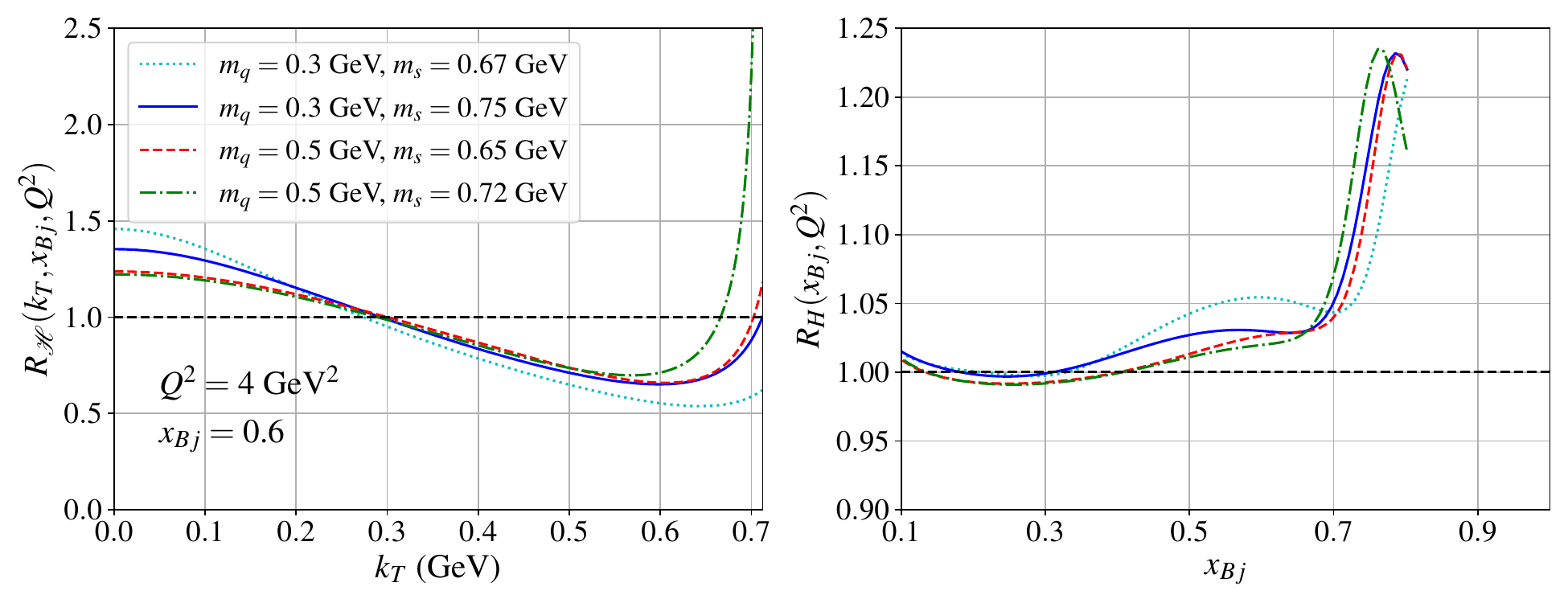}}
  \caption{(Left) The ratio between the entropy $k_T$-integrand of Eq.~(\ref{eq:entropy}) as computed for the exact relative to the factorized $k_T$-unintegrated structure function in the spectator model of Ref.~\cite{Moffat:2017sha}; namely, we evaluate $R_\mathcal{H} = \mathcal{H}^\mathrm{exact}_{F_1}(x, k_T, Q^2) / \mathcal{H}^\mathrm{fact}_{F_1}(x, k_T, Q^2)$. (Right) The analogue of this quantity in the $k_T$-integrated case --- {\it i.e.}, the differential entropy ratio, $R_\mathit{H} = \mathit{H}^\mathrm{exact}_{F_1}(x, Q^2) / \mathit{H}^\mathrm{fact}_{F_1}(x, Q^2)$; both panels are evaluated at $Q^2 = 4\,\mathrm{GeV}^2$ and with the model parameter assumptions indicated in the left-panel legend.
  }
  \label{fig-2}
\end{figure}

In Fig.~\ref{fig-2} (left), we plot the ratio of the entropy $k_T$-integrand from Eq.~(\ref{eq:entropy}) within the exact calculation, which explicitly involves factorization-breaking contributions, relative to the factorized case; similarly, in the right panel, we plot the corresponding ratio of the $k_T$-integrated differential entropy. We note that, owing to the nature of the continuous distributions on which the differential entropies are evaluated, the entropy itself is free to become negative, particularly in regions of greater localization of the system in configuration space. As such, the ratios shown in Fig.~\ref{fig-2} represents relative variations in the {\it magnitude} of the entropies.

Inspecting Fig.~\ref{fig-2} (left), we see that the inclusion of factorization-breaking effects in the {\it exact} model calculation drives a relative increase in the $k_T$-dependent contribution to the entropy (making it more sharply negative), while suppressing it at larger values of $k_T$. In the right panel we find that the relative differential entropy begins to diverge as $x_{Bj}$ gets larger, signifying the growing role of factorization-breaking effects as $W$ decreases toward the region of shallow-inelastic scattering.

\section{Conclusion}
\label{sec:conc}
The results highlighted above indicate a connection between the entropy of distributions associated with interacting QCD bound states and the quantum mechanically meaningful localization of the system. While these results will be discussed further in a forthcoming study~\cite{2025QuantumEntropy}, they suggest the possibility of valuable insights which might flow from further development of theoretical relations among definitions of the entropy, properties of quantum correlation functions and associated amplitudes, and quantum mechanical notions of entanglement and decoherence.
The refinement of this formalism could open the opportunity to extract additional information from future quantum simulation(s) of QCD bound states and processes; for example, calculable entropies which might be directly evaluated in quantum simulations~\cite{Khor:2023xar} may provide independent information on the formation of structure or interpretation of dynamics which break fundamental QCD relations like factorization theorems.

\section*{Acknowledgments}
The work of Henry Bloss was supported by the U.S.~Department of Energy, Office of Science, Office of Workforce Development for Teachers and Scientists (WDTS) under the Science Undergraduate Laboratory Internship (SULI) program.
The work of Brandon Kriesten and T. J. Hobbs at Argonne National Laboratory was supported by the U.S.~Department of Energy under contract DE-AC02-06CH11357.

\end{document}